\documentclass[11pt]{article}
\usepackage{amsmath,amssymb,color,graphics,epsfig,cite}
\usepackage{graphicx,subfigure}
\usepackage{amsfonts}
\newcommand{\hoch}[1]{$\, ^{#1}$}

\newcommand{\be}{\begin{equation}}
	\newcommand{\ee}{\end{equation}}
\newcommand{\bea}{\setlength\arraycolsep{2pt} \begin{eqnarray}}
	\newcommand{\eea}{\end{eqnarray}}
\newcommand{\nn}{\nonumber}

\def\crampest{\medmuskip = 1mu plus 1mu minus 1mu}

\def\ft#1#2{{\textstyle{\frac{\scriptstyle #1}{\scriptstyle #2} } }}
\def\fft#1#2{{\frac{#1}{#2}}}

\def\0{{\sst{(0)}}}
\def\1{{\sst{(1)}}}
\def\2{{\sst{(2)}}}
\def\3{{\sst{(3)}}}
\def\4{{\sst{(4)}}}
\def\5{{\sst{(5)}}}
\def\6{{\sst{(6)}}}
\def\7{{\sst{(7)}}}
\def\8{{\sst{(8)}}}
\def\sst#1{{\scriptscriptstyle #1}}

\begin{document}
	
\begin{center}

{\Large {\bf Exact Toda Black Holes of Rank-2 Lie Groups}}
		
\vspace{20pt}

H. L\"u\hoch{1,2}, Peng-Yu Wu\hoch{1}, Ze-Hua Wu\hoch{1} and Weicheng Zhao\hoch{1}

\vspace{10pt}

{\it \hoch{1}Center for Joint Quantum Studies, Department of Physics,\\
			School of Science, Tianjin University, Tianjin 300350, China }

\medskip

{\it \hoch{2}The International Joint Institute of Tianjin University, Fuzhou,\\ Tianjin University, Tianjin 300072, China}

\vspace{40pt}
		
\underline{ABSTRACT}

\end{center}

We consider Einstein gravity coupled to two Maxwell fields and one dilatonic scalar, and construct spherically-symmetric and static black holes that are charged under both Maxwell fields in general $D$ dimensions. We find that for suitable dilaton couplings, the equations of motion can be cast into one-dimensional Toda equations of all rank-2 Lie groups. We devise a brute-force approach to obtain the most general but remarkably elegant solutions to the Toda equations. This allows us to construct exact black holes associated with all the rank-2 Lie groups. We study their thermodynamics and verify explicitly an earlier claim in the literature that all these thermodynamic quantities can be derived without having to solve for these black hole solutions.

\vfill {\footnotesize mrhonglu@gmail.com\ \ \ wupy2023@tju.edu.cn\ \ \  wuzh\_2022@tju.edu.cn\ \ \  zhaoweichengok@tju.edu.cn}
	
	\thispagestyle{empty}
	\pagebreak
	\tableofcontents
	\addtocontents{toc}{\protect\setcounter{tocdepth}{2}}
	
	\newpage

\section{Introduction}
\label{sec:intro}

High nonlinearity is one of the most outstanding characteristics of Einstein's General Relativity (GR), compared to the quantum field theories describing the other three observed fundamental interactions. Despite this nonlinearity, exact solutions can be constructed in GR. (See \cite{Stephani:2003tm} for a summary of known exact solutions.)  Among these, the Schwarzschild and Kerr black holes are arguably the most important vacuum solutions. Their solvability can be understood simply as a linear perturbation of the Minkowski vacuum in a suitable coordinate choice, e.g.~the Kerr-Schild ansatz \cite{Kerr:1965wfc, Bini:2010hrs}. The Kerr-Schild form belongs to the algebraic special class of solutions when the matter energy-momentum tensor satisfies the null energy condition. A more general class of exact solutions based on algebraic speciality can be constructed \cite{Stephani:2003tm}. (See recent new examples in \cite{Lu:2025fzm}.)

When matter is involved, the most notable example is the Reissner-Nordstr\"om (RN) black hole of Einstein-Maxwell gravity. There are two ways to understand the solvability of the RN black hole. In the Schwarzschild-like coordinates, the solution can also be easily cast into the Kerr-Schild form and the full solution arises simply as a linear perturbation. In the $p$-brane coordinate of string theory research, where the transverse space is Euclidean, the equations of motion can be cast into a one-dimensional Liouville equation, which is integrable. In particular, in the extremal limit, the metric is governed by a harmonic function on the three-dimensional Euclidean transverse space. This second approach can be generalised to a general class of Einstein-Maxwell-Dilaton (EMD) theory. Indeed, the equations of motion of spherically-symmetric and static charged black holes in EMD theory can be cast into a Liouville equation which can be solved completely \cite{Lu:1996hh}. The general solutions are more than black holes, involving three independent parameters, the mass $M$, electric charge $Q$ and a scalar charge $\Sigma$. The three-parameter solutions typically have naked singularities, a consequence of violating the black hole no-scalar hair conjecture. For the solution to describe a black hole, the scalar charge $\Sigma$ has to be fine-tuned appropriately so that it becomes a specific function of the mass and charge, namely $\Sigma=\Sigma(M,Q)$ \cite{Lu:1996hh}. The electrically-charged EMD black holes were constructed much earlier and are now known as GMGHS black holes \cite{Gibbons:1987ps,Garfinkle:1990qj}.

Liouville equation is a special case of more general integrable Toda equations associated with classical Lie groups. If one can generalize the EMD theory so that the black hole equations of motion can be reduced to the Toda equations of certain Lie groups, then exact solutions of black holes naturally emerge. Indeed, the equations of motion of the dyonic black hole in the Kaluza-Klein theory in four dimensions can be cast into the $SL(3,\mathbb R)$ Toda equations, and hence can be solved exactly \cite{Gibbons:1985ac}. (See \cite{Rasheed:1995zv,Gimon:2007mh} for the rotating version and \cite{Lu:2013ura} for the anti-de Sitter (AdS) generalization.) In \cite{Lu:2013uia}, Einstein gravity minimally coupled to a set of Maxwell fields and dilaton scalars were constructed such that the electrically-charged black holes are governed by a set of $A_n\sim SL(n+1,\mathbb R)$ Toda equations, and corresponding exact black hole solutions were straightforwardly constructed \cite{Lu:2013uia}. (See also \cite{Lu:1996jr,Lu:1997rd,Ivashchuk:1999jd,Ivashchuk:2002ge} for a variety of Toda $p$-brane or cosmological solutions constructed using the supergravity scalar cosets.)

Generally speaking, for a Lie group of rank $n$, the EMD theory should be generalized to involve $n$ Maxwell fields and $(n-1)$ dilatonic scalars $\vec \phi=(\phi_1, \phi_2, \ldots, \phi_{n-1})$, with their non-minimum couplings in the form of
\be
-\ft14 e^{\vec a_i \cdot \vec \phi} F_i^2\,,\qquad i=1,2,\ldots, n\,.
\ee
The reduction of the equations of motion to Toda equations requires some specific choice of the dilaton coupling constants $\vec a_i$. In particular, the required dilaton coupling constants for the $A_n$ Toda equations were given in \cite{Lu:2013uia}.

There has been significant progress in constructing Toda solutions systematically in the mathematical community \cite{Olshanetsky:1978km,Kostant:1979qu,Farwell:1982du, Farwell:1983sx}. However, the solutions are generally complicated, except for the $A_n$ case, where an elegant presentation of the solutions exists \cite{Anderson:1995sz}. These solutions were utilized to construct general $A_n$ Toda black holes \cite{Lu:2013uia}. (See also some non-systematic but explicit examples in \cite{Lu:1996jr,Lu:1997rd}.) In this paper, we focus on constructing Toda black holes associated with rank-2 Lie groups, which consists of $D_2\sim A_1\times A_1$, $A_2$, $B_2\sim C_2$ and $G_2$ inequivalent examples. The associated gravitational theory is Einstein-Maxwell-Maxwell-Dilaton (EMMD) gravity, involving one dilaton and two Maxwell fields. In particular, the $D_2$ and $A_2$ black holes were known in literature. In this paper, we present two new exact black holes, associated with the $B_2$ and $G_2$ Toda equations, which can also be obtained from the Toda flux-brane construction \cite{Ivashchuk:2013jja}.

The general procedure of the construction of solutions to Toda equations including the $B_2$ and $G_2$ groups can be found in \cite{Olshanetsky:1978km,Kostant:1979qu, Farwell:1982du,Farwell:1983sx}. However, the resulting solutions appear to be enormously complex; consequently, the black hole solution would be so complicated that it is rendered effectively useless; it would not be much better than  numerical solutions.  Further complication arises from the fact that the most general Toda equations involve four independent integration constants, parametrizing the mass $M$, scalar charge $\Sigma$ and two electric charges $(Q_1^e,Q_2^e)$. However, it can be easily demonstrated that black holes emerge only after we fine-tune the scalar charge as a specific function of the mass and two electric charges. From a complicated general Toda solution, this fine-tuning is not easy to perform, even with an exact solution. In this paper, based on the mathematical fact that the Toda equations of $B_2$ and $G_2$ are solvable, we devise a brute-force approach and obtain the most general and remarkably elegant solutions to the Toda equations. The simplicity of our Toda solutions makes the fine-tuning of the scalar charge simple, which then allows to construct all the Toda black holes associated with all the rank-2 Lie groups.

The paper is organized as follows. In section 2, we present the EMMD theory and its equations of motion. We consider the spherically-symmetric and static ansatz, and illustrate that a black hole involves three parameters, the mass and two electric charges. The scalar charge must be fine-tuned to be a certain specific function of the aforementioned three parameters. We present a convenient ansatz that allows us to cast the equations of motion to a set of Toda-like equations. We determine the appropriate dilaton coupling constants that lead to the Toda equations of each rank-2 group. In section \ref{sec:toda}, we devise a brute-force method to solve all the Toda equations to obtain general solutions with four independent parameters. We discuss how to fine-tune these parameters so that general solutions with naked singularity reduce to the three-parameter black holes. In section \ref{sec:b2g2}, we present exact $B_2$ and $G_2$ Toda black holes and analyse their global structure by deriving the complete set of thermodynamic quantities that satisfy the first law of black hole thermodynamics. In section \ref{sec:a3b2}, we review the general $SL(n+1,\mathbb R)$ Toda black hole, and illustrate that the $B_2$ Toda black hole can be obtained from the truncation of the $A_3$ Toda black hole.

This work is also motivated by recent works on the EMMD theories in \cite{Lu:2025eub,Yang:2025rud}, where it was demonstrated that black hole thermodynamic quantities can be derived without having to solve for the equations of motion. The works of \cite{Lu:2025eub,Yang:2025rud} focused on $D=4$ dimensions. In section \ref{sec:thermo}, we generalize the statements to general $D$ dimensions, and use the Toda black holes to confirm these statements. We then present further conjectures for more general theories with multiple Maxwell and dilaton fields. We conclude the paper in section \ref{sec:con}. In Appendix \ref{app:review}, we present the previously-known $D_2$ and $A_2$ black holes in general dimensions. In Appendix \ref{app:extremal}, we give extremal $B_2$ and $G_2$ Toda black holes in two explicit charges.

\section{EMMD black holes and Toda equations}
\label{sec:emmd}

\subsection{Theory and ansatz}

The field content of the EMMD theory in general $D$ dimensions consists of the metric, two Maxwell fields $A_i$ and a dilatonic (massless) scalar $\phi$. The Lagrangian is given by
\be
{\cal L} = \sqrt{-g} \Big( R - \ft12 (\partial\phi)^2 - \ft14 e^{a_1\phi} F_1^2 -
\ft14 e^{a_2\phi} F_2^2\Big),\qquad F_i=dA_i\,,\label{EMMD}
\ee
where $(a_1,a_2)$ are two constants, typically referred to as the dilaton coupling constants. The Lagrangian is inspired by string theory. In particular, when the dilaton coupling constants take the values $(a_1,a_2)=(\sqrt3,-1/\sqrt3), (1,-1), (\sqrt3,-\sqrt3)$, the EMMD theory can be embedded in STU supergravity \cite{Duff:1995sm}. It was also noted that for some EMMD theories, the near-horizon geometry has conformal AdS$_6$ or AdS$_7$ structures that can be embedded in string theory \cite{Bu:2025nci}.
The equations of motion associated with the variation of the dilaton, the Maxwell fields and the metric are respectively given by
\bea
\Box\phi &=& \ft14 a_1 e^{a_1\phi} F_1^2 + \ft14 a_2 e^{a_2\phi} F_2^2\,,\qquad
\nabla_\mu \big(e^{a_1 \phi} F_1^{\mu\nu}\big)=0=\nabla_\mu \big(e^{a_2 \phi} F_2^{\mu\nu}\big)\,,\nn\\
G_{\mu\nu}&=& \ft12 \big(\partial_\mu\phi \partial_\nu \phi -\ft12 g_{\mu\nu} (\partial\phi)^2\big) +\ft12 e^{a_1\phi} \big((F_1^2)_{\mu\nu} - \ft14 g_{\mu\nu} F_1^2\big) + \ft12 e^{a_2\phi} \big((F_2^2)_{\mu\nu} - \ft14 g_{\mu\nu} F_2^2\big)\,.
\label{covariant-eom}
\eea
Note that, when $a_2=-a_1$, the equations of motion of the spacetimes electrically charged under both Maxwell fields are mathematically equivalent to those of dyonic spacetime of a single Maxwell field.

In this paper, we consider spherically-symmetric and static spacetime configurations that are electrically charged under both Maxwell fields. The most general ansatz takes
the form
\be
ds^2 = - h(r) dt^2 + \fft{dr^2}{f(r)} + r^2 d\Omega_{D-2}^2\,,\qquad
A_i= \psi_i(r) dt\,,\qquad \phi=\phi(r)\,.\label{ansatz-0}
\ee
The first integration of the Maxwell field can be easily solved, giving rise to two integration constants $(Q_1,Q_2)$:
\be
F_1 = \fft{4Q_1}{r^{D-2}} e^{-a_1 \phi} \sqrt{\fft{h}{f}}\, dt\wedge dr\,,\qquad
F_2 = \fft{4Q_2}{r^{D-2}} e^{-a_2 \phi} \sqrt{\fft{h}{f}}\, dt\wedge dr\,.
\ee
The two electric charges are then given by
\be
Q^e_i = \fft{1}{16\pi} \int e^{a_i \phi} {*F_i} = \fft{\Omega_{D-2}}{4\pi} Q_i\,,\qquad\hbox{where}
\qquad \Omega_{n} = \frac{2 \pi^{\frac{n+1}{2}}}{\Gamma(\frac{n+1}{2})}.\label{echarges}
\ee
Note that $\Omega_n$ is the volume of a unit round $n$-sphere. Note that $(Q_1^e, Q_2^e)$ are the two electric charges, whilst $(Q_1,Q_2)$ are referred to as the electric charge parameters. In four dimensions, we simply have $(Q_1^e, Q_2^e) = (Q_1,Q_2)$.

In this paper, we shall focus only on asymptotically-flat spacetimes. The most general solution involves two more parameters, the mass $M$ and scalar charge $\Sigma$, defined from the asymptotic falloffs as
\be
g_{tt}\sim  -1 + \fft{16\pi}{(D-2)\Omega_{D-2}} \fft{M}{r^{D-3}} + \cdots\,,\qquad
\phi = \fft{16\pi}{(D-3) \Omega_{D-2}} \fft{\Sigma}{r^{D-3}} + \cdots\,.
\ee
Here we have made a convention that the speed of light is $c=1$ in the asymptotic Minkowski spacetime. We further set $\phi(\infty)=0$, by making use of the constant dilaton shifting symmetry of the Lagrangian, compensated by some appropriate constant scaling of the Maxwell fields. The $1/r^{D-3}$ power falloff is the signature condensation of the massless fields, i.e. the graviton and dilaton. The dilaton constant shifting symmetry implies that there is a conserved charge \cite{Lu:2025eub,Yang:2025rud}, which we define to be
\be
\Sigma=\fft{1}{16\pi} \int_{r\rightarrow \infty} {\tilde *d\phi}\,,
\ee
where $\tilde *$ is the Hodge dual defined in the $(D-1)$-dimensional subspace of constant time slice of \eqref{ansatz-0}, namely
\be
d\tilde s_{D-1}^2 = \fft{dr^2}{f(r)} + r^2 d\Omega_{D-2}^2\,.
\ee
Thus, the most general solution involves four parameters, $(M, \Sigma, Q_1^e, Q_2^e)$.

In the asymptotic infinity region, the theory becomes a quantum field theory of free scalar and Maxwell fields. The long-range force between two identical spacetimes reduces to standard Newton's and Coulomb's laws, given by
\be
-r^{D-3} \hbox{force}\Big|_{r\rightarrow \infty} = M^2 + \fft{2(D-2)}{D-3}\Big( \Sigma^2 - (Q_1^e)^2 -(Q_2^e)^2\Big)\,.\label{longrangeforce}
\ee
Gravity and scalar force are attractive while the electric forces of two identical particles are repulsive. This generalizes the four-dimensional works of \cite{Cremonini:2021upd,Cremonini:2022sxf,Cremonini:2023vwf} and \cite{Lu:2025eub,Yang:2025rud}.

However, it is important to note that the general solution with four independent parameters $(M,\Sigma, Q_1,Q_2)$ cannot describe a black hole, but instead a solution with a naked singularity. This can be seen from the scalar equation, which can be cast into the following form, if a black hole exists with the event horizon located at $r_+$, namely
\be
\int_{r_+}^{\infty} \Big(r^{D-2} \sqrt{hf} \phi'^2 - \fft{8\phi}{r^{D-2}}\sqrt{\ft{h}{f}} \big(a_1 Q_1^2 e^{-a_1\phi} + a_2 Q_2^2 e^{a_2 \phi}\big)\Big) dr =0\,.
\ee
This constraint is not automatically satisfied by the equations of motion, and it requires a fine-tuning of the four parameters. Specifically, in the black hole solution, the scalar charge is not independent, but a specific function of the mass and electric charges, i.e.
\be
\Sigma = \Sigma(M, Q_1^e, Q_2^e)\,.\label{sigmamq1q2}
\ee
For black holes with exact solutions, such a relation can be straightforwardly obtained. However, when exact solutions are not known, determining such a fine-tuning can be an involved numerical task.

Note that the long-range force law \eqref{longrangeforce} emphasizes the role of the scalar charge, which is typically ignored in many discussions of black holes, such as black hole thermodynamics. As we shall see later, the scalar charge plays an important role in deriving the black hole thermodynamics without solving for the solutions, even though it does not enter the first law directly and is typically not regarded as a thermodynamic quantity.

\subsection{Toda and Toda-like equations}

Since the construction of GMGHS solutions, there have been significant progress in constructing black hole or $p$-brane solutions in the EMMD-type of theories \eqref{EMMD}. Many useful formulae can be found in \cite{Lu:1995cs}. Solving equations directly using the ansatz \eqref{ansatz-0} at this later stage of research is not the most effective approach. Inspired by the works of \cite{Geng:2018jck,Cribiori:2022cho} and \cite{Cremonini:2021upd,Cremonini:2023vwf}, we propose the ansatz
\bea
ds^2 &=& - H_1^{-\fft{2\delta}{\delta+a_1^2}} H_2^{-\fft{2\delta}{\delta+a_2^2}} f dt^2 +
H_1^{\fft{2\delta}{(D-3)(\delta+a_1^2)}} H_2^{\fft{2\delta}{(D-3)(\delta+a_2^2)}} \Big(\fft{dr^2}{f} + r^2 d\Omega_{D-2}^2\Big)\,,\nn\\
F_1 &=&  \fft{4Q_1}{r^{D-2} H_1^2} H_2^{-\fft{2(\delta+a_1 a_2)}{\delta+a_2^2}} dt\wedge dr\,,\qquad
F_2 = \fft{4Q_2}{r^{D-2} H_2^2} H_1^{-\fft{2(\delta+a_1 a_2)}{\delta+a_1^2}} dt\wedge dr\,,\nn\\
\phi &=& \fft{2a_1}{\delta + a_1^2} \log H_1 + \fft{2 a_2}{\delta + a_2^2} \log H_2\,,
\label{ansatz-1}
\eea
where $(H_1,H_2,f)$ are all functions of the radial coordinate $r$. This ansatz is inspired by single-charge configurations where analytic solutions exist for general dilaton constants; it can be viewed as a certain nonlinear ``superposition'' of two singly-charged Ans\"atze.

Substituting the ansatz \eqref{ansatz-1} into the equations of motion \eqref{covariant-eom}, the two Maxwell equations are automatically satisfied. The remaining equations are actually consistent for arbitrary $\delta$, with
\be
f=1 - \fft{2\mu}{r^{D-3}} + \fft{Q^2}{r^{2(D-3)}}\,.
\ee
where $(\mu,Q)$ are two integration constants. However, while it will ultimately lead to the correct solution, the ansatz \eqref{ansatz-1} with the above $f$ and arbitrary $\delta$ is not the most convenient one to adopt. To simplify the ansatz further, we note that the EMMD theory reduces to an EMD theory provided that we have either $Q_1=0$ or $Q_2=0$. The electrically charged black holes of the EMD theory are well understood, and they are described by $f=1-2\mu/r^{D-3}$, together with
\be
Q_2=0\,,\qquad H_2=1\,,\qquad  H_1\sim 1 + \fft{Q_1}{r^{D-3}}\,,
\ee
or with the subscripts ``1'' and ``2'' interchanged. Thus, we need to further impose $H_i=1$ if $Q_i=0$. This leads to
\be
f=1 - \fft{2\mu}{r^{D-3}}\,,\qquad \delta=\fft{2(D-3)}{D-2}\,.\label{fdelta}
\ee
When $D=4$, the ansatz \eqref{ansatz-1} then reduces to the one proposed in \cite{Lu:2025eub,Yang:2025rud}. It is perhaps worth observing that the parameter $\delta$ also appears in the Smarr relation in the Schwarzschild black hole as $M = \fft{2}{\delta} T S$. It is rather mysterious that this particular $\delta$ also appears in the ansatz \eqref{ansatz-1}, and its significance is not at all apparent.

We find that all the equations of motion now reduce to two second-order differential equations for functions $(H_1, H_2)$:
\bea
\frac{H_1''}{H_1} -\frac{H_1'^2}{H_1^2}+\frac{(D-3)+f}{r f} \fft{H_1'}{H_1} &=&- \frac{4 \left(\delta+a_1^2\right) Q_1^2}{ r^{2(D-2)} f} H_1^{-2} H_2^{-\frac{2 \left(\delta+a_1 a_2\right)}{\delta+a_2^2}}\,,\nn\\
\frac{H_2''}{H_2} -\frac{H_2'^2}{H_2^2}+\frac{(D-3)+f}{r f} \fft{H_2'}{H_2} &=&- \frac{4 \left(\delta+a_2^2\right) Q_2^2}{ r^{2(D-2)} f} H_2^{-2} H_1^{-\frac{2 \left(\delta+a_1 a_2\right)}{\delta+a_1^2}}\,,\label{H12pp}
\eea
together with the (consistent) first-order Hamiltonian constraint
\bea
0&=& \fft{1}{\delta+a_1^2} \fft{H_1'^2}{H_1^2} + \fft{1}{\delta+a_2^2} \fft{H_2'^2}{H_2^2} + \frac{2 \left(\delta+a_1 a_2\right)}{\left(\delta+a_1^2\right) \left(\delta+a_2^2\right)} \fft{H_1'H_2'}{H_1 H_2}\nn\\
&&
-\fft{2(D-3)\mu}{r^{D-2}f} \Big( \fft{1}{\delta+a_1^2} \fft{H_1'}{H_1} + \fft{1}{\delta+a_2^2} \fft{H_2'}{H_2}\Big)\nn\\
&&-\fft{4}{r^{2(D-2)}f}\Big(Q_2^2\, H_2^{-2} H_1^{-\frac{2 \left(\delta+a_1 a_2\right)}{\delta+a_1^2}}+Q_1^2\, H_1^{-2} H_2^{-\frac{2 \left(\delta+a_1 a_2\right)}{\delta+a_2^2}}\Big).\label{hamcons}
\eea
Here a prime denotes a derivative with respect to $r$. Thus, we have a consistent set of equations that guarantee solutions. In the asymptotic Minkowski region, we can choose, without loss of generality, that $H_1(\infty)=1=H_2(\infty)$, which implies that $\phi(\infty)=0$. The leading falloffs are of $1/r^{D-3}$ order, giving rise to two independent parameters, the mass $M$ and scalar charge $\Sigma$, defined by
\bea
&&g_{tt}  \sim - 1 + \fft{2m}{r^{D-3}} + \cdots \,,\qquad \phi \sim \fft{2\sigma}{r^{D-3}} + \cdots \,,\nn\\
&& M=\fft{(D-2)\Omega_{D-2}}{8\pi} m\,,\qquad
\Sigma=\fft{(D-3)}{8\pi} \Omega_{D-2}\,\sigma\,.
\eea
Together with the two electric charges $(Q_1^e,Q_2^e)$, the most general asymptotically-flat solutions involve a total of four independent parameters $(M,\Sigma, Q_1^e,Q_2^e)$. As we shall see later, the most general solution does not describe a black hole, which arises only when the scalar charge $\Sigma$ is fine-tuned to be a suitable function of the three conserved quantities $(M,Q_1^e,Q_2^e)$. Black holes emerge when the functions $(H_1,H_2)$ run smoothly from certain nonzero finite quantities on the horizon $r=r_+$ to asymptotic $r\rightarrow \infty$, where
\be
r_+=(2\mu)^{\fft{1}{D-3}}\,.\label{rplus}
\ee
In other words, the event horizon is located at $f(r_+)=0$. Employing the standard technique, we can derive the temperature and entropy, given by
\be
T=\fft{D-3}{4 \pi r_+ } H_1(r_+)^{-\fft{2}{\delta+a_1^2}}H_2(r_+)^{-\fft{2}{\delta+a_2^2}}\,,\qquad
S=\ft14\Omega_{D-2} r_+^{D-2}H_1(r_+)^{\fft{2}{\delta+a_1^2}} H_2(r_+)^{\fft{2}{\delta+a_2^2}}\,.\label{TandS}
\ee
It is important to note, for later purposes, that the product of temperature and entropy is proportional to $\mu$. Specifically, we have
\be
T S=\fft{(D-3)\Omega_{D-2}}{8 \pi}\mu\,.\label{TSproduct}
\ee
The $D=4$ result was known in \cite{Lu:2025eub,Yang:2025rud}. One way to test the validity of the black hole solution is to verify the first law of black hole thermodynamics, namely
\be
dM=TdS + \Phi_1 dQ_1^e + \Phi_2 dQ_2^e\,.\label{firstlaw}
\ee
Intriguingly, as we shall see later, the long-range force \eqref{longrangeforce} between two of these identical  black holes all reduces to the identity
\be
M^2 +\fft{4}{\delta} \Big(\Sigma^2 -(Q_1^e)^2 +(Q_2^e)^2\Big) =
\fft{(D-2)^2\Omega_{D-2}^2 }{64 \pi^2}\mu^2\,.\label{forcelaw}
\ee
This is valid even when the solution is not a black hole, with an independent scalar charge. A proof in $D=4$ was given in \cite{Lu:2025eub}, and the proof in general dimensions follows straightforwardly.

In order to understand better the nature of the equations \eqref{H12pp} and \eqref{hamcons}, we need to massage them further. We first make a field redefinition and introduce $(q_1,q_2)$ functions:
\be
H_1 = \lambda_1  f^{\fft{a_2(\delta+a_1^2)}{2\delta(a_2-a_1)}}\, e^{-q_1} \,,\qquad
H_2 = \lambda_2  f^{\fft{a_1(\delta+a_2^2)}{2\delta(a_1-a_2)}}\, e^{-q_2}\,,\label{h1h2redef}
\ee
with the constants $(\lambda_1, \lambda_2)$ related to the charge parameters by
\be
Q_1^2 = \fft{(D-3)^2}{4 (\delta+a_1^2)} \lambda_1^2\, \lambda_2^{\fft{2(\delta+a_1a_2)}{\delta+a_2^2}}\,,\qquad
Q_2^2 = \fft{(D-3)^2}{4 (\delta	+a_2^2)}\lambda_2^2\,\lambda_1^{\fft{2(\delta+a_1a_2)}{\delta+a_1^2}}\,.
\ee
We then make a coordinate transformation and introduce a new radial coordinate $\eta$, defined by
\be
e^{-2\mu \eta} = f\,.\label{ctransf}
\ee
The asymptotic flat region then corresponds to $\eta=0$. If the spacetime describes a black hole, it follows from \eqref{rplus} that the event horizon is located at $\eta=\infty$.

The equations in \eqref{H12pp} can now be cast into one-dimensional Toda-like equations
\be
\ddot q_i = \exp({\sum_{j=1}^2 K_{ij} q_j})\,,\qquad i=1,2\,,\qquad\hbox{with}\qquad
K=\left(
    \begin{array}{cc}
      2 & \alpha_+ \\
      \alpha_- & 2 \\
    \end{array}
  \right).\label{todalikeeom}
\ee
Here, a dot denotes a derivative with respect to $\eta$. The $2\times 2$ matrix $K$ is determined solely by dilaton coupling constants $(a_1,a_2)$, with
\be
\alpha_+= \fft{2(\delta + a_1 a_2)}{\delta+a_2^2}\,,\qquad
\alpha_-=\fft{2(\delta + a_1 a_2)}{\delta+a_1^2}\,.
\ee
The Hamiltonian constraint \eqref{hamcons}  becomes
\be
\fft{\delta}{\delta + a_1^2} (\dot q_1^2-\ddot q_1) + \fft{\delta}{\delta + a_2^2} (\dot q_2^2-\ddot q_2) +
\fft{2 (\delta + a_1 a_2) \delta}{(\delta + a_1^2)(\delta +a_2^2)} \dot q_1\dot q_2
= \mu^2.\label{q1q2cons}
\ee
It is understood that $\ddot q_i$ in the above equation should be replaced by the right-hand-side exponential expressions in \eqref{todalikeeom}. It is easy to verify that this quadratic first-order Hamilton constraint is consistent with the second-order differential equations \eqref{todalikeeom}. In fact, we can read off the effective Lagrangian $L=T-V$ from \eqref{q1q2cons}, where the kinetic and potential energies are respectively given by
\bea
T &=& \fft{\delta}{\delta + a_1^2} \dot q_1^2 + \fft{\delta}{\delta + a_2^2} \dot q_2^2 +
\fft{2 (\delta + a_1 a_2) \delta}{(\delta + a_1^2)(\delta +a_2^2)} \dot q_1\dot q_2\,,\nn\\
V &=& -\fft{\delta}{\delta+a_1^2} e^{2q_1 + \alpha_+ q_2} -
 \fft{\delta}{\delta+a_2^2} e^{2q_2 + \alpha_- q_1}\,.
\eea
Thus, we see that the two-charge system of the EMMD theory reduces to the Toda-like two-particle mechanics.
The fact that the potential energy can be expressed as acceleration, as in \eqref{q1q2cons}, is very special to the exponential type of energy. It is easy to verify that the equations of motion in \eqref{todalikeeom} now simply arise from the Euler-Lagrange equations, and the Hamiltonian constraint \eqref{q1q2cons} simply means the ``conservation of energy'': $T+V=\mu^2$. We refer to this as the gravitational constraint, as $\mu$ originates from the ansatz \eqref{ansatz-1}.

When the matrix $K$ is a Cartan matrix of a certain Lie group, then the system of equations becomes that of one-dimensional Toda equations of the corresponding group. In classical Lie groups, there are four inequivalent rank-2 Lie groups, namely $D_2$, $A_2$, $B_2\sim C_2$ and $G_2$. ($D_2\sim A_1\times A_1$ is reducible.) Their Cartan matrices and corresponding dilaton couplings are summarized in Table \ref{table1}. As we can see, all the Toda equations correspond to having $a_1a_2<0$, which implies that the dilaton scalar can be truncated out if $a_1 F_1^2 + a_2 F_2^2=0$. The theory then reduces to Einstein-Maxwell gravity. In particular, when the two electric charges satisfy $Q_1=\sqrt{-a_2/a_1} Q_2$, the two-charge black hole reduces to the RN black hole. For later purposes,  the field redefinition \eqref{h1h2redef} can now be expressed as
\be
H_1=\lambda_1 e^{-\fft{2(2-\alpha_+)}{4-\alpha_+\alpha_-} \mu \eta} e^{-q_1}\,,\qquad
H_2=\lambda_2 e^{-\fft{2(2-\alpha_-)}{4-\alpha_-\alpha_+} \mu \eta} e^{-q_2}\,.\label{H12eta}
\ee
The regularity on the horizon implies that $(H_1,H_2)$ must be finite and nonzero as $\eta\rightarrow \infty$.

\begin{table}
\begin{center}
\begin{tabular}{|c|c|c|c|c|}
  \hline
  % after \\: \hline or \cline{col1-col2} \cline{col3-col4} ...
   & $\alpha_+$ & $\alpha_-$ & $a_1$ & $a_2$ \\
   \hline
  $D_2$ & 0 & 0 & any & $-\delta/a_1$ \\
  \hline
 $A_2$ & -1 & -1 & $\sqrt{3\delta}$ & $-\sqrt{3\delta}$ \\
  \hline
  $B_2$ & -2 & -1 & $3\sqrt{\delta}$ & $-2\sqrt{\delta}$ \\
  \hline
 $G_2$ & -3 & -1 & $3\sqrt{3\delta}$ & $-5\sqrt{\delta/3}$ \\
  \hline
\end{tabular}
\caption{\small The Cartan matrices of the rank-2 Lie groups and the corresponding dilaton coupling constants. The parameter $\delta$ is given by \eqref{fdelta}, with $\delta=1$ in four dimensions.}\label{table1}
\end{center}
\end{table}

The Toda equations are integrable and hence the most general solutions associated with the spherically-symmetric and static ansatz \eqref{ansatz-1} with \eqref{fdelta} can be obtained. Before presenting the solutions, we examine the number of nontrivial integration constants. The two second-order Toda equations have a total of four integration constants, corresponding precisely to $(M,\Sigma,Q_1^e, Q_2^e)$. Specifically, it follows from \eqref{H12eta} that $H_1=1=H_2$ at the asymptotic infinity ($\eta=0$) relates two integration constants of the Toda equations to $(Q_1^e, Q_2^e)$. The Hamiltonian constraint $T+V=\mu^2$ relates $\mu$ to a certain quadratic combination of the integration constants. The fourth one corresponds to the scalar charge, which becomes a function of mass and charges if we impose regularity of \eqref{H12eta} on the horizon ($\eta= \infty$.)

\section{Elegant solutions from a brute-force approach}
\label{sec:toda}

There have been great successes in solving the Toda equations for general Lie groups \cite{Olshanetsky:1978km,Kostant:1979qu,Farwell:1982du, Farwell:1983sx}. For the rank-2 systems we study in this paper, we adopt a new brute-force approach, since computing power is readily available. We make an ansatz
\be
e^{-q_1} = \sum_{i,j=-N}^N a_{ij} e^{i \eta_1 + j \eta_2}\,,
\qquad
e^{-q_2} = \sum_{i,j=-N}^N b_{ij} e^{i \eta_1 + j \eta_2}\,,
\ee
where $N$ is a certain finite integer and
\be
\eta_1=\mu_1 \eta + c_1\,,\qquad \eta_2 = \mu_2 \eta + c_2\,.
\ee
Here we introduce four integration constants $(\mu_1,\mu_2,c_1,c_2)$. For a generic set of Toda-like equations \eqref{todalikeeom}, the integer $N$ will be infinite even when $\alpha_\pm$ are integers. In other words, there is no closed form and the series involves an infinite number of terms. However, for the integrable Toda systems, the equations can be solved when number $N$ is finite. In fact, for the rank-2 groups, we find that the $N$ is a small number. Specifically, for $D_2,A_2$ and $B_2$, we have $N=1$. For $G_2$, we have $N=2$. Substituting the ansatz into the Toda equations, we have rational polynomials of $(z_1,z_2)=(e^{\eta_1}, e^{\eta_2})$, with coefficients constructed from constants $(a_{ij}, c_{ij}, \mu_1,\mu_2,c_1,c_2)$. The original Toda differential equations therefore reduce to a set of algebraic polynomial equations for these constants. We refer to this method as a ``brute force'' not because it requires heavy computation. In fact, the algebraic polynomial equations are easy and quick to solve. We call it a brute-force method simply because it does not require us to know the intricate knowledge of the relevant groups. Nevertheless, using this method, we obtain a set of the most general and yet remarkably elegant solutions to the Toda equations.

\bigskip
\noindent{\bf $D_2$ Toda solutions}. The $D_2\sim A_1\times A_1$ is reducible and the Toda equations are simply two copies of the Liouville equations:
\be
\ddot q_1 = e^{2q_1}\,,\qquad \ddot q_2 = e^{2q_2}\,.\label{2liouville}
\ee
They are subject to the first-order Hamiltonian constraint, associated with our gravitational ansatz:
\be
\fft{\dot q_1^2 - \ddot q_1}{1+a_1^2}
+\fft{\dot q_2^2 - \ddot q_2}{1 + a_2^2} =\mu^2\,.
\ee
The most general solution to \eqref{2liouville} is
\be
e^{-q_i} = \fft{1}{\mu_i} \sinh\eta_i\,,\qquad i=1,2.
\ee
The solution contains four independent integration constants $(\mu_1,\mu_2,c_1,c_2)$. Related to our spherically-symmetric and static gravitational ansatz \eqref{ansatz-1}, the parameters $(\mu_1,\mu_2)$ are subject to the gravitational constraint
\be
\fft{\mu_1^2}{1+a_1^2} + \fft{\mu_2^2}{1+a_2^2}=\mu^2\,.
\ee
The constants $(c_1,c_2)$ are on the other hand free at this stage.

\bigskip
\noindent{\bf $A_2$ Toda solutions.} The $A_2$ Toda equations and the first-order Hamiltonian constraint are given by
\bea
&&\ddot q_1 = e^{2 q_1 - q_2}\,,\qquad \ddot q_2=e^{2q_2 - q_1}\,,\nn\\
&&\ft14 (\dot q_1^2 -\ddot q_1) + \ft14 (\dot q_2^2 - \ddot q_2) -
\ft14 \dot q_1\dot q_2=\mu^2\,.
\eea
The most general solution is
{\crampest
\bea
\ft12(e^{-q_1} + e^{-q_2}) &=& \frac{\cosh \eta_1}{\left(\mu _1-\mu _2\right) \left(2 \mu _1+\mu _2\right)}+\frac{\cosh \eta_2}{\left(\mu _1-\mu _2\right) \left(\mu _1+2 \mu _2\right)} -\frac{\cosh (\eta_1+\eta_2)}{\left(2 \mu _1+\mu _2\right) \left(\mu _1+2 \mu _2\right)}\,,\nn\\
\ft12(e^{-q_1} - e^{-q_2}) &=& \frac{\sinh \eta_1}{\left(\mu _1-\mu _2\right) \left(2 \mu _1+\mu _2\right)}+\frac{\sinh \eta_2}{\left(\mu _1-\mu _2\right) \left(\mu _1+2 \mu _2\right)} +\frac{\sinh (\eta_1+\eta_2)}{\left(2 \mu _1+\mu _2\right) \left(\mu _1+2 \mu _2\right)}\,,
\eea}
together with the gravitational constraint
\be
\ft14 (\mu_1^2 + \mu_1\mu_2 + \mu_2^2) = \mu^2\,.
\ee
This solution is a special case of the $SL(n,\mathbb R)$ Toda solutions obtained in \cite{Lu:2013uia}.

\bigskip
\noindent{\bf $B_2$ Toda solutions.} The $B_2$ Toda equations are
\be
\ddot q_1 = e^{2 q_1 - 2 q_2}\,,\qquad
\ddot q_2 = e^{2 q_2 - q_1}\,,
\ee
with the gravitational constraint
\be
\ft1{10}\big(\dot q_1^2 - \ddot q_1\big) +
\ft15\big(\dot q_2^2 - \ddot q_2\big) - \ft15 \dot q_1 \dot q_2 = \mu^2\,.
\ee
Using our brute-force approach, we find that the most general solution can be elegantly expressed as
\bea
e^{-q_1} &=& \fft{1}{(\mu_1^2-\mu_2^2)^2}\bigg(2 + 2\cosh\eta_1 \cosh\eta_2
-\Big(\fft{\mu_1}{\mu_2} + \fft{\mu_2}{\mu_1}\Big) \sinh\eta_1 \sinh\eta_2\bigg)\,,\nn\\
e^{-q_2} &=& \fft{1}{\mu_1^2-\mu_2^2} \Big(
\fft{1}{\mu_1} \sinh\eta_1 + \fft{1}{\mu_2}\sinh\eta_2\Big)\,,
\eea
subject to the gravitational constraint
\be
\ft1{10} (\mu_1^2 + \mu_2^2) = \mu^2\,.
\ee

\bigskip
\noindent{\bf $G_2$ Toda solutions.} The equations are
\be
\ddot q_1 = e^{2 q_1 - 3 q_2}\,,\qquad
\ddot q_2 = e^{2 q_2 - q_1}\,,
\ee
together with the Hamiltonian constraint relevant to gravitational ansatz \eqref{ansatz-1}
\be
\ft1{28} (\dot q_1^2 - \ddot q_1) + \ft3{28} (\dot q_2^2 - \ddot q_2) - \ft3{28} \dot q_1 \dot q_2=\mu^2\,.
\ee
We find that the most general solution is
\bea
e^{-q_1} &=& \fft{2}{\mu _1^2\mu _2^2 \left(2 \mu _1-\mu _2\right) \left(2 \mu _2-\mu _1\right) \left(\mu _1^2-\mu _2^2\right)^2}  \Bigg(\frac{12 \left(\mu _1^2-\mu _2 \mu _1+\mu _2^2\right)}{\left(\mu _1-2 \mu _2\right) \left(2 \mu _1-\mu _2\right)}\nn\\
&&-\frac{3 \left(\mu _1-2 \mu _2\right) \left(\mu _1+\mu _2\right) \cosh \eta_1 }{\left(\mu _1-\mu _2\right) \mu _2} -\frac{3 \left(2 \mu _1-\mu _2\right) \left(\mu _1+\mu _2\right) \cosh \eta_2 }{\mu _1 \left(\mu _1-\mu _2\right)} \nn\\
&&+\frac{\left(\mu _1-\mu _2\right)^2 \cosh (\eta_1 + \eta_2)}{\mu _1 \mu _2}
+\frac{3 \left(\mu _1+\mu _2\right)^2 \cosh(\eta_1-\eta_2)}{\mu _1 \mu _2}
\nn\\
&&-\frac{\mu _1^2 \left(\mu _1+\mu _2\right) \cosh(\eta_1-2\eta_2)}{\left(\mu _1-2 \mu _2\right) \left(\mu _1-\mu _2\right) \mu _2} -\frac{\mu _2^2 \left(\mu _1+\mu _2\right) \cosh (2\eta_1-\eta_2)}{\mu _1 \left(\mu _1-\mu _2\right)
\left(2 \mu _1-\mu _2\right)}\Bigg),\nn\\
%%%%%%%%
e^{-q_2} &=& \frac{2}{\mu _1^2\mu_2^2 \left(\mu _1^2-\mu _2^2\right)}
\bigg(\frac{\mu _1 \cosh\eta _2}{\mu _1-2 \mu _2}+\frac{\mu _2 \cosh\eta_1}{2 \mu _1-\mu _2}-\frac{\mu _1+\mu _2}{\mu _1-\mu _2}\nn\\
&&-\frac{\mu _1 \mu _2 \left(\mu _1+\mu _2\right) \cosh (\eta _1-\eta_2)}{\left(\mu _1-2 \mu _2\right) \left(\mu _1-\mu _2\right) \left(2 \mu _1-\mu _2\right)} \bigg),
\eea
subject to the gravitational constraint
\be
\ft1{28} (\mu_1^2 + \mu_2^2 - \mu_1\mu_2) = \mu^2\,.
\ee

{\noindent \bf Application in black hole constructions.} By construction, in all our general Toda solutions, there are four integration constants, namely $(\mu_1,\mu_2,c_1,c_2)$. In our original gravitational ansatz \eqref{ansatz-1}, we have already introduced three constants $(\mu,Q_1,Q_2)$ that parametrize the mass and two electric charges. It appears that the whole system would have a total of seven parameters. This is in fact not the case. The Hamiltonian constraint gives a quadratic relation between $(\mu_1,\mu_2)$ and $\mu$. The requirement that $H_i(\infty)=1$ determines the constants $(c_1,c_2)$ in terms of $(\mu_1,\mu_2,Q_1,Q_2)$. Thus the most general spherically-symmetric and static solution of \eqref{ansatz-1} contains the four integration constants $(\mu_1,\mu_2,Q_1,Q_2)$, parametrizing the four independent hairy parameters, $(M,\Sigma, Q_1,Q_2)$.

However, as we have mentioned earlier, A generic solution with four independent parameters does not describe a black hole. We now discuss the conditions that lead to a black hole geometry on and outside of the horizon. The coordinate transformation \eqref{ctransf} implies that the coordinate ranges of $r$ and $\eta$ on and outside of the horizon are
\be
r\in [r_+, \infty) \qquad \Leftrightarrow\qquad \eta\in[\infty,0)\,.
\ee
It follows from \eqref{H12eta} that whether $H_1$ and $H_2$ are regular on the horizon ($\eta=\infty$) depends on the values $(\mu_1,\mu_2)$. For generic values that are subject to the Hamiltonian constraint, these functions are singular on the would-be horizon $\eta=\infty$ and hence the solution does not describe a black hole. For suitable choices, the functions $(H_1,H_2)$ become regular at $\eta=\infty$, giving rise to well-defined black hole solutions. It follows from \eqref{H12eta} that we have
\bea
D_2:&&\qquad \mu_1=\mu\,,\qquad \mu_2=\mu\,,\nn\\
A_2:&&\qquad \mu_1=\mu\,,\qquad \mu_2=-\mu\,,\nn\\
B_2:&&\qquad \mu_1=3\mu\,,\qquad \mu_2=\mu\,,\nn\\
G_2:&&\qquad \mu_1=2\mu\,,\qquad \mu_2=-4\mu\,.
\eea
These choices are not unique, but different choices to black holes are equivalent.
Having made these choices, we find that the functions $(H_1,H_2)$ take the form
\be
H_1=\sum_{i=0}^{N_1} \beta_i f^i\,,\qquad
H_2=\sum_{i=0}^{N_2} \gamma_i f^i\,,
\ee
for some finite integer values of $(N_1,N_2)$. Thus, $(H_1,H_2)$ may run smoothly from the horizon ($f(r_+)=0$) to the asymptotic infinity with $f(\infty)=1$. Therefore, the black hole solutions all have three independent parameters, namely $(\mu,Q_1,Q_2)$, parametrizing the mass and two electric charges.  We shall present the explicit black hole solutions in the next two sections.

\section{$B_2$ and $G_2$ Toda black holes}
\label{sec:b2g2}

In the previous section, we obtained the most general and exact solutions to the Toda equations of all rank-2 Lie groups, and discussed how the black hole solutions emerge by fine-tuning of the four integration constants. The $D_2$ and $A_2$ Toda black holes were known in the literature and we shall present them in the appendix \ref{app:review}.  Here, we give the exact $B_2$ and $G_2$ Toda black holes and analyse their thermodynamic properties.

\subsection{$B_2$ Toda black holes}

The $B_2$ black hole arises as the dilaton coupling constants are $a_1=3\sqrt{\delta}$ and $a_2=-2\sqrt{\delta}$. The most general two-charge non-extremal black hole is actually quite simple, given by \eqref{ansatz-0} with
\bea
H_1 &=&\gamma_1^{-1}\Big(1 -4 \beta _1 f  +6 \beta _1 \beta _2 f^2 -4 \beta _1 \beta _2^2 f^3+ \beta _1^2 \beta _2^2 f^4\Big),\nn\\
H_2 &=&\gamma_2^{-1}\Big( 1-3 \beta _2 f + 3\beta _1 \beta _2 f^2 -\beta _1 \beta _2^2 f^3\Big),
\eea
where $(\gamma_1,\gamma_2)$ are polynomials of constants $(\beta_1, \beta_2)$:
\be
\gamma_1 = 1 - 4\beta_1 + 6 \beta_1\beta_2 - 4 \beta_1 \beta_2^2 + \beta_1^2 \beta_2^2\,,\qquad
\gamma_2 =1-3\beta_2 + 3\beta_1\beta_2 - \beta_1\beta_2^2\,.
\ee
Note that the coefficients are precisely the binomial coefficients such that when $\beta_1=\beta_2$, the polynomials factorize to become binomials of certain order. The horizon is located at $r_+$ with $f(r_+)=0$, so that we have $H_i (r_+) = 1/\gamma_i$. The asymptotic region corresponds to $r\rightarrow \infty$, with $f=1$ and $H_i=1$. Note that instead of deriving the black hole solution from the general Toda solution, the black hole can also be obtained from the Toda flux-brane construction by directly replacing the $z=1/r^{D-3}$ in \cite{Ivashchuk:2013jja} by the function $f$, with an appropriate overall scalings so that $H_i$ approaches unity at $r\rightarrow \infty$. The same technique applies to the $G_2$ case as well.

The general black hole solution contains three independent parameters, namely $(\mu, \beta_1, \beta_2)$, parametrizing the mass and two electric charges. Specifically, the electric charges are
\be
Q_1^e=\frac{\Omega_{D-2}\sqrt{(D-3)(D-2)\,\beta_1} \gamma _2}{4\pi\sqrt5\,\gamma _1} \mu \,,\qquad
Q_2^e=\frac{\Omega_{D-2}\sqrt{3(D-3)(D-2)\beta_2\gamma_1}}{4\pi\sqrt{10}\,\gamma _2} \mu\,.
\ee
The corresponding electric potentials are respectively given by
\be
\Phi_1=2\sqrt{\fft{2\beta_1}{5\delta}} \gamma_2^{-1} (1 - 3\beta_2+3\beta_2^2- \beta_1\beta_2^2)\,,\qquad
\Phi_2 = 2\sqrt{\fft{3\beta_2}{5\delta\gamma_1}} (1 -2\beta_1 + \beta_1 \beta_2)\,.
\ee
The mass and scalar charge are given by
\bea
M &=& \frac{(D-2) \Omega _{D-2}}{40 \pi  \gamma _1 \gamma _2}
\Big(5 -16 \beta _1-9 \beta _2 + 45 \beta _1 \beta _2 -25 \beta _1 \beta _2^2
-25 \beta _1^2 \beta _2^2 +45 \beta _1^2 \beta _2^3\nn\\
&&-9 \beta _1^3 \beta _2^3 -16 \beta _1^2 \beta _2^4 + 5 \beta _1^3 \beta _2^4\Big)\mu\,,\nn\\
\Sigma &=& \frac{3 \sqrt{2(D-3)(D-2)} \Omega _{D-2}}{20 \pi  \gamma _1 \gamma _2}  \left(\beta _1-\beta _2\right)(1 - 5\beta_1\beta_2 + 5 \beta_1\beta_2^2 -\beta_1^2\beta_2^3)\mu\,.\label{masssig}
\eea
The general formulae for temperature and entropy are given by \eqref{TandS}. Specifically, they are given by
\be
T=\fft{D-3}{4\pi r_+} (\gamma_1 \gamma_2^2)^{\fft{1}{5\delta}}\,,\qquad
S=\ft14 \Omega_{D-2}\, r_+^{D-2} (\gamma_1 \gamma_2^2)^{-\fft{1}{5\delta}}\,.
\ee
With these, we find that the long-range force identity \eqref{forcelaw} is satisfied. Furthermore, the first law \eqref{firstlaw} can be easily verified. Note that the scalar charge is also a function of $(\mu, \beta_1,\beta_2)$; it is not an independent parameter and can be expressed in terms of the mass and two electric charges. The scalar charge $\Sigma$ does not enter the first law, and hence it is not considered a thermodynamic variable; however, it plays an important role in the long-range force. As we shall see later, it also is instrumental to derive the all the thermodynamic variables without needing the solution. It follows from \eqref{masssig} that when $\beta_2=\beta_1$, the scalar charge vanishes. Furthermore, we have $Q_1/Q_2=\sqrt{2/3}$ and the solution reduces to the RN black hole.

The extremal limit corresponds to taking the blackening parameter $\mu\rightarrow 0$ limit, while keeping the electric charges $(Q_1^e, Q_2^e)$ finite and nonzero. This implies that we need to take the parameter $\beta_1$ and $\beta_2$ to 1. Specifically, we first introduce two parameters $(c_1,c_2)$ and take
\be
\beta_1=1-\frac{2 c_1 \mu }{c_2^2}\,,\qquad
\beta_2=1-\frac{2 c_1 \mu }{c_2^2}-\frac{4 c_1 \left(c_2^2-c_1^2\right) \mu ^3}{c_2^6}\,,
\ee
then we substitute the above into the solution and finally take $\mu \rightarrow 0$. We find
\bea
H_1 &=& 1+\frac{2  \left(3 c_1^2-c_2^2\right) c_2^2}{c_1 \left(3 c_1^2-2 c_2^2\right)} \rho +\frac{6 c_2^4}{3 c_1^2-2 c_2^2} \rho ^2 +\frac{4 c_2^6}{c_1 \left(3 c_1^2-2 c_2^2\right)} \rho ^3+\frac{ c_2^8}{c_1^2 \left(3 c_1^2-2 c_2^2\right)}\rho ^4 \,,\nn\\
H_2 &=& 1+3 c_1 \rho+3  c_2^2 \rho ^2+\frac{ c_2^4}{c_1} \rho ^3\,,\qquad \rho=\fft{1}{r^{D-3}}\,.
\eea
The temperature of the extremal black hole is zero by construction. The remaining thermodynamic quantities are
\bea
Q_1^e &=& \frac{c_2^4 \sqrt{(D-3) (D-2)} \Omega _{D-2}}{8 \sqrt{5} \pi  c_1 \left(3 c_1^2-2 c_2^2\right)}\,,\qquad Q_2^e =
\fft{\Omega_{D-2}}{8\pi} \sqrt{\ft3{10} (D-3) (D-2) \left(3 c_1^2-2 c_2^2\right)}\,,\nn\\
M_{\rm ext} &=& \frac{\left(9 c_1^4-3 c_2^2 c_1^2-c_2^4\right) (D-2) \Omega _{D-2}}{40 \pi  c_1 \left(3 c_1^2-2 c_2^2\right)}\,,\qquad \Phi_1^{\rm ext} = \frac{\sqrt{2} \left(3 c_1^2-c_2^2\right)}{\sqrt{5\delta}\, c_2^2 }\,,\nn\\
S_{\rm ext} &=& \ft14 \Omega _{D-2}^{\frac{1}{3-D}} \Big(\frac{320\, 2^{3/5} \pi ^2
}{3^{3/5} (D-3) (D-2)} (Q^e_1)^{\fft4{5}} (Q_2^e)^{\fft6{5}}\Big)^{\fft{1}{\delta}}\,,\qquad \Phi_2^{\rm ext} = \frac{2 \sqrt{3} c_1}{\sqrt{5\delta\,(3 c_1^2-2 c_2^2)}}\,.
\eea
The scalar charge in the extremal limit is
\be
\Sigma_{\rm ext} = -\frac{3 \sqrt{2 (D-3)(D-2)} \Omega _{D-2} \left(c_1^2-c_2^2\right) \left(6 c_1^2-c_2^2\right)}{80 \pi  c_1 \left(3 c_1^2-2 c_2^2\right)}\,.
\ee
It is easy to verify that the long-range force between the two identical extremal black holes vanishes. The solution reduces to the extremal RN black hole when $c_1=c_2$. In appendix \ref{app:extremal}, we present the extremal solution using explicitly the electric charges.

\subsection{$G_2$ Toda black holes}

The $G_2$ black hole arises when the dilaton coupling constants are $a_1=3\sqrt{3\delta}$ and $a_2=-5\sqrt{\delta/3}$. The most general two-charge black hole solution is given by \eqref{ansatz-0}, with
\bea
H_1 &=& \gamma_1^{-1}\Big(1 - 10 \beta_1 f + 45 \beta _1 \beta _2 f^2-120 \beta _1 \beta _2^2 f^3 +15 \beta _1 \beta _2^2 \left(5 \beta _1+9 \beta _2\right) f^4 -252 \beta _1^2 \beta _2^3 f^5\nn\\
&&+15 \beta _1^2 \beta _2^3 \left(9 \beta _1+5 \beta _2\right) f^6 -120 \beta _1^3 \beta _2^4 f^7+45 \beta _1^3 \beta _2^5 f^8-10 \beta _1^3 \beta _2^6 f^9+
\beta _1^4 \beta _2^6 f^{10}\Big)\,,\nn\\
H_2 &=&\gamma^{-1}_2 \Big(1 - 6\beta_2 f + 15 \beta _1 \beta _2 f^2 -20 \beta _1 \beta _2^2 f^3+15 \beta _1 \beta _2^3 f^4 -6 \beta _1^2 \beta _2^3 f^5 + \beta _1^2 \beta _2^4 f^6\Big)\,,
\eea
where
\bea
\gamma_1 &=& 1 - 10 \beta_1 + 45 \beta _1 \beta _2-120 \beta _1 \beta _2^2 +15 \beta _1 \beta _2^2 \left(5 \beta _1+9 \beta _2\right) -252 \beta _1^2 \beta _2^3\nn\\
&&+15 \beta _1^2 \beta _2^3 \left(9 \beta _1+5 \beta _2\right) -120 \beta _1^3 \beta _2^4 f^7+45 \beta _1^3 \beta _2^5-10 \beta _1^3 \beta _2^6+
\beta _1^4 \beta _2^6\,,\nn\\
\gamma_2 &=& 1 - 6\beta_2 + 15 \beta _1 \beta _2 -20 \beta _1 \beta _2^2+15 \beta _1 \beta _2^3 -6 \beta _1^2 \beta _2^3 + \beta _1^2 \beta _2^4\,.
\eea
The solution contains three integration constants $(\mu,\beta_1,\beta_2)$, parametrizing the mass and two electric charges. The electric charge parameters are given by
\be
Q_1^e=\fft{\Omega_{D-2} \,\mu}{8 \pi  \gamma_1}\sqrt{\ft57 (D-3)(D-2) \beta_1 \gamma_2^3}\,,\qquad
Q_2^e=\fft{3\Omega_{D-2}\, \mu}{8 \pi \gamma_2} \sqrt{\ft17 (D-3)(D-2) \beta_2 \gamma_1} \,.
\ee
Their electric potentials are
\bea
\Phi_1&=&\sqrt{\fft{10\beta_1}{7\delta\gamma_2^3}}\Big(
1-9 \beta _2+36 \beta _2^2-30 \beta _1 \beta _2^2-54 \beta _2^3+126 \beta _1 \beta _2^3-81 \beta _1^2 \beta _2^3\nn\\
&&-45 \beta _1 \beta _2^4+84 \beta _1^2 \beta _2^4-36 \beta _1^2 \beta _2^5+9 \beta _1^2 \beta _2^6-\beta _1^3 \beta _2^6\Big),\nn\\
\Phi_2 &=& 3\sqrt{\fft{2\beta_2}{7\delta \gamma_1}}\Big(1-5 \beta _1+10 \beta _1 \beta _2-10 \beta _1 \beta _2^2+5 \beta _1^2 \beta _2^2-\beta _1^2 \beta _2^3\Big).
\eea
The mass and scalar charge are
\bea
M &=& \frac{(D-2)\Omega _{D-2}}{56 \pi  \gamma _1 \gamma _2} \mu \Big(7-65 \beta _1-33 \beta _2+630 \beta _1 \beta _2-525 \beta _1^2 \beta _2-1925 \beta _1 \beta _2^2\nn\\
&&+2800 \beta _1^2 \beta _2^2+3570 \beta _1 \beta _2^3-8631 \beta _1^2 \beta _2^3+2730 \beta _1^3 \beta _2^3-2835 \beta _1 \beta _2^4+11284 \beta _1^2 \beta _2^4\nn\\
&&-8435 \beta _1^3 \beta _2^4+2025 \beta _1^4 \beta _2^4-1575 \beta _1^2 \beta _2^5+1575 \beta _1^4 \beta _2^5-2025 \beta _1^2 \beta _2^6+8435 \beta _1^3 \beta _2^6\nn\\
&&-11284 \beta _1^4 \beta _2^6+2835 \beta _1^5 \beta _2^6-2730 \beta _1^3 \beta _2^7+8631 \beta _1^4 \beta _2^7-3570 \beta _1^5 \beta _2^7-2800 \beta _1^4 \beta _2^8\nn\\
&&+1925 \beta _1^5 \beta _2^8+525 \beta _1^4 \beta _2^9-630 \beta _1^5 \beta _2^9+33 \beta _1^6 \beta _2^9+65 \beta _1^5 \beta _2^{10}-7 \beta _1^6 \beta _2^{10}
\Big),\nn\\
%%%%%
\Sigma &=& \frac{15 \sqrt{6 (D-3)(D-2)} \Omega _{D-2}}{112 \pi  \gamma _1 \gamma _2} \mu (\beta _1-\beta _2)\Big(1-35 \beta _1 \beta _2+140 \beta _1 \beta _2^2-189 \beta _1 \beta _2^3-112 \beta _1^2 \beta _2^3\nn\\
&&+735 \beta _1^2 \beta _2^4-540 \beta _1^3 \beta _2^4-540 \beta _1^2 \beta _2^5+735 \beta _1^3 \beta _2^5-112 \beta _1^3 \beta _2^6-189 \beta _1^4 \beta _2^6\nn\\
&&+140 \beta _1^4 \beta _2^7-35 \beta _1^4 \beta _2^8+\beta _1^5 \beta _2^9
\Big).\label{masssigg2}
\eea
The general formulae \eqref{TandS} for temperature and entropy can now be evaluated to become
\be
T=\fft{(D-3)}{4\pi r_+} (\gamma_1\gamma_2^3)^{\fft1{14\delta}}\,,\qquad
S=\fft14 \Omega_{D-2} r_+^{D-2} (\gamma_1\gamma_2^3)^{-\fft1{14\delta}}\,.
\ee
With these, we find that the long-range force identity \eqref{forcelaw} is satisfied. Furthermore, the first law \eqref{firstlaw} can be easily verified. It follows from \eqref{masssigg2} that the scalar charge vanishes when $\beta_2=\beta_1$, in which case, we have $Q_1/Q_2=\sqrt{5/9}$ and a resulting RN black hole.

To take the extremal limit, we introduce $(c_1,c_2)$ parameters and define
\be
\beta_1=1-\frac{2 c_1^2 }{c_2^3} \mu\,,\qquad
\beta_2=1-\frac{2 c_1^2 }{c_2^3} \mu+\frac{16 c_1^4 \left(c_1^6-c_2^6\right) }{9 c_2^{15}} \mu ^5\,.
\ee
Substitute these into the black hole solution, and take $\mu\rightarrow 0$ limit, we have
\bea
H_1 &=& 1 +\fft{1}{c^{12}}\Big(30 c_1^4 c_2^3 \left(5 c_1^6-3 c_2^6\right) \rho + 90 c_1^2 c_2^6 \left(5 c_1^6-2 c_2^6\right) \rho ^2
-180 c_2^9 \left(c_2^6-5 c_1^6\right) \rho ^3\nn\\
 &&+ \frac{90 c_2^{12} \left(15 c_1^6-c_2^6\right)}{c_1^2} \rho ^4 + \frac{18 c_2^{15} \left(85 c_1^6-c_2^6\right) }{c_1^4} \rho ^5
+1260 c_2^{18} \rho ^6 +\frac{720 c_2^{21}}{c_1^2} \rho ^7\nn\\
 &&+\frac{270 c_2^{24}}{c_1^4} \rho ^8+\frac{60  c_2^{27}}{c_1^6} \rho ^9 +\frac{6 c_2^{30}}{c_1^8} \rho ^{10}\Big),\qquad \rho=\fft{1}{r^{D-3}}\,,\nn\\
H_2 &=& 1 +\frac{\left(5 c_1^6+c_2^6\right) }{c_1^2 c_2^3}  \rho +15  c_1^2 \rho ^2 +20 c_2^3 \rho ^3 +\frac{15  c_2^6}{c_1^2} \rho ^4 +\frac{6  c_2^9}{c_1^4} \rho ^5+\frac{ c_2^{12}}{c_1^6} \rho ^6\,,
 \eea
where $c^{12}= 25 c_1^{12}-20 c_2^6 c_1^6+c_2^{12}$. The temperature vanishes in the extremal limit. The remaining thermodynamic quantities and the scalar charge are given by
\bea
(Q_1^e,Q_2^e) &=&  \fft{\sqrt{(D-3)(D-2)}\Omega_{D-2}}{16\pi}
\left(\frac{6 \sqrt{\frac{5}{7}} c_1 c_2^{12}}{c^{12}},\frac{\sqrt{\frac{3}{14}} c^6}{c_1^2 c_2^3}\right),\nn\\
(\Phi_1,\Phi_2) &=&
\left(\frac{c_1^3 \left(5 c_1^6-3 c_2^6\right) \sqrt{\frac{5}{14 \delta }}}{c_2^9},\frac{\left(5 c_1^6+c_2^6\right) \sqrt{\frac{3}{7 \delta }}}{c^6}\right)\,,\nn\\
M &=& \frac{3 \left(125 c_1^{18}-25 c_2^6 c_1^{12}-45 c_2^{12} c_1^6+c_2^{18}\right) (D-2) \Omega _{D-2}}{224 \pi  c^{12} c_1^2 c_2^3}\,,\nn\\
\Sigma &=& -\frac{5 \sqrt{6 (D-3)(D-2)} \left(c_1^6-c_2^6\right) \left(125 c_1^{12}-40 c_2^6 c_1^6-c_2^{12}\right)  \Omega _{D-2}}{448 \pi  c^{12} c_1^2 c_2^3}\,,\nn\\
S &=& \ft14 \Omega _{D-2}^{\frac{1}{3-D}} \Big(\frac{1792 \pi ^2 (Q^e_1)^{5/7} (Q^e_2)^{9/7}}{3\ 3^{2/7} 5^{5/14} (D-3) (D-2)}\Big)^{\fft1{\delta}}\,.
\eea
In appendix \ref{app:extremal}, we present the extremal solution in terms of explicit electric charges.

\section{From $A_3$ to $B_2$ Toda black holes}
\label{sec:a3b2}

It is well known that the non-simply-laced $B_n$ can be embedded into a larger simply-laced $D_{n+1}$. This embedding can be easily seen from the appropriate folding of the Dynkin diagram of the simple-laced $D_{n+1}$ algebras. For $n=2$, we have $D_3$, which is isomorphic to $A_3$. Thus the EMMD theory for the $B_2$ Toda black hole we constructed should arise from a consistent construction of the theory that gives rise to $A_3$
Toda black hole. The general exact $A_{n}$ Toda black holes were constructed in \cite{Lu:2013uia}. For our purpose, we consider the $A_3$ example. The Lagrangian involves three Maxwell fields, together with two dilatonic scalars:
\begin{equation}
{\cal L}= \sqrt{-g} \Big(R - \fft12 {\partial_\mu \vec\phi} \cdot \partial^\mu \vec \phi - \fft14 \sum_{i=1}^3 e^{\vec a_i \cdot \vec \phi}\, F_i^2\Big)\,,\label{sl4r}
\end{equation}
where $\vec\phi=(\phi_1,\phi_2)$. The dilaton coupling constant vectors $\vec a_i$ satisfy the dot product \cite{Lu:2013uia}
\be
\vec a_1 \cdot \vec a_1=\vec a_2 \cdot \vec a_2=\vec a_3 \cdot \vec a_3= 9 \delta\,,\qquad
 \vec a_1\cdot \vec a_2=\vec a_2 \cdot \vec a_3 = -6 \delta\,,\qquad
\vec a_1\cdot \vec a_3 = -\delta\,.\label{sl4raiaj}
\ee
Note that the three $\vec a_i$ vectors are spanned in two-dimensional space and cannot be linearly independent.  It follows from (\ref{sl4raiaj}) that we have
\begin{equation}
3 \vec a_1 + 4 \vec a_2 + 3 \vec a_3=0\,.
\end{equation}
One way to find a specific representation of the dilaton vectors is
\be
\vec a_1 = (-2, \sqrt{5}) \sqrt{\delta}\,,\qquad \vec a_2 = (3,0) \sqrt{\delta}\,,\qquad \vec a_3 = (-2,-\sqrt5) \sqrt{\delta}\,.
\ee
With this choice of representation, we find that it is consistent to set $F_1=F_3$ and
$\phi_2=0$,  which precisely leads to the EMMD theory associated with the $B_2$ case.

The $A_3$ Toda black hole is given by \cite{Lu:2013uia}
\begin{eqnarray}
ds^2 &=& - (H_1 H_2 H_3)^{-\fft15} f dt^2 + (H_1 H_2 H_3)^{\fft{1}{5}} (f^{-1} dt^2 + r^2 d\Omega_{D-2}^2)\,,\cr
\vec \phi &=& \fft{1}{5\delta} \Big( \vec a_1 \log H_1 + \vec a_2 \log H_2 + \vec a_3 \log H_3\Big)\,,\qquad f=1-\fft{2\mu}{r^{D-3}}\,,\cr
A_1 &=& \sqrt{\ft{3(D-2)}{5(D-3)}}\, \fft{1-2 \beta_1 f + \beta_1\beta_2 f^2}{\sqrt{\beta_1\gamma_2}\, H_1}\,dt\cr
A_2 &=& \sqrt{\ft{4(D-2)}{5(D-3)}}\,\fft{1-3\beta_2 f + \ft32 \beta_2(\beta_1 + \beta_2) f^2 -\beta_1\beta_2\beta_3 f^3}{\sqrt{\beta_2\gamma_1\gamma_3}\, H_2}\,dt\,,\cr
A_3 &=&  \sqrt{\ft{3(D-2)}{5(D-3)}}\, \fft{1-2\beta_3 f+ \beta_2\beta_3 f^2}{\sqrt{\beta_3\gamma_2}\,H_3}\,dt\,,
\end{eqnarray}
where
\begin{eqnarray}
H_1 &=& \gamma_1^{-1} (1 - 3 \beta_1 f + 3 \beta_1 \beta_2 f^2 - \beta_1 \beta_2 \beta_3 f^3)\,,\cr
H_2 &=& \gamma_2^{-1} (1 - 4 \beta_2 f + 3 \beta_2 (\beta_1 + \beta_3) f^2 - 4 \beta_1 \beta_2 \beta_3 f^3 +
 \beta_1 \beta_2^2 \beta_3 f^4)\,,\cr
H_3&=& \gamma_3^{-1} (1 - 3 \beta_3 f + 3 \beta_2 \beta_3 f^2 - \beta_1 \beta_2 \beta_3 f^3)\,,
\end{eqnarray}
with
\begin{eqnarray}
\gamma_1 &=&1 - 3 \beta_1 + 3 \beta_1 \beta_2 - \beta_1 \beta_2 \beta_3\,,\cr
\gamma_2 &=&1 - 4 \beta_2 + 3 \beta_2(\beta_1 + \beta_3) - 4 \beta_1 \beta_2 \beta_3 + \beta_1 \beta_2^2 \beta_3\,, \cr
\gamma_3 &=&1 - 3 \beta_3 + 3 \beta_2 \beta_3 -
 \beta_1 \beta_2 \beta_3\,.
 \end{eqnarray}
We can turn off any charge by setting the associated $\beta_i$ to zero.  When all the $\beta_i$ parameters are equal, the solution reduces to the RN black hole. Setting $\beta_3=\beta_1$ leads to $F_3=F_1$, in which case, we have indeed $\phi_2=0$. We thus obtain the exact $B_2$ Toda black hole from the $A_3$ Toda black hole.

For later purposes, we also present all thermodynamical quantities of the $A_3$ Toda black hole, obtained in \cite{Lu:2013uia}. After correcting a few typos in \cite{Lu:2013uia}, we have the thermodynamic quantities:
\bea
M&=&\fft{(D-2)\Omega_{D-2}\,\mu}{8\pi} \Big(1 + \fft{3\beta_1}{5\gamma_1}(1-2\beta_2+\beta_2\beta_3)\cr
&&+ \fft{4\beta_2}{5\gamma_2}(1 - \ft32(\beta_1+\beta_3) + 3\beta_1\beta_3 - \beta_1\beta_2\beta_3)+ \fft{3\beta_3}{5\gamma_3} (1 - 2 \beta_2 + \beta_1 \beta_2)\Big)\,,\cr
Q_1^e&=&\fft{\Omega_{D-2}\,\mu}{8\pi \gamma_1}\sqrt{\ft35(D-2)(D-3)\beta_1\gamma_2}\,,\cr
Q_2^e&=&\fft{\Omega_{D-2}\,\mu}{8\pi \gamma_2}\sqrt{\ft45(D-2)(D-3)\beta_2\gamma_1\gamma_3}\,,\cr
Q_3^e &=&\fft{\Omega_{D-2}\,\mu}{8\pi \gamma_3}\sqrt{\ft35(D-2)(D-3)\beta_3\gamma_2}\,,\nn\\
\Phi_1&=&\sqrt{\ft{3(D-2)\beta_1}{5(D-3)\gamma_2}}\, (1-2\beta_2 + \beta_2\beta_3)\,,\cr
\Phi_2&=&\sqrt{\ft{4(D-2)\beta_2}{5(D-3)\gamma_1\gamma_3}}\, (1-
\ft32(\beta_1 + \beta_3) + 3\beta_1\beta_3 -\beta_1\beta_2\beta_3)\,,\cr
\Phi_3&=&\sqrt{\ft{3(D-2)\beta_3}{5(D-3)\gamma_2}}\, (1-2\beta_2 + \beta_1\beta_2)\,,\nn\\
T &=& \ft{D-3}{4\pi r_+} (\gamma_1 \gamma_2 \gamma_3)^{\fft{D-2}{10(D-3)}}\,, \qquad
S=\ft14 \Omega_{D-2} \Big(\ft{(2\mu)^{10}}{\gamma_1 \gamma_2 \gamma_3} \Big)^{ \fft{D-2}{10(D-3)}}\,.
\eea
It is straightforward to verify the first law of thermodynamics
\begin{equation}
dM=T dS + \Phi_1 dQ_1^e + \Phi_2 dQ_2^e + \Phi_3 dQ_3^e\,.
\end{equation}
An important ingredient that was absent in \cite{Lu:2013uia}, but will be important for our later discussion is the scalar charge, namely
\be
\vec \Sigma = \fft{1}{16\pi} \int_{r\rightarrow \infty} {{\tilde *} d\vec \phi} \equiv (\Sigma_1,
\Sigma_2)\,.
\ee
For the $A_3$ black hole, we find
\bea
\Sigma_1 &=&\frac{3 \sqrt{(D-3) (D-2)} \mu  \Omega _{D-2}}{20 \sqrt{2} \pi }\Big(-\frac{\beta _1 \left(\beta _3 \beta _2-2 \beta _2+1\right)}{\gamma _1}-\frac{\beta _3\left(\beta _1 \beta _2-2 \beta _2+1\right) }{\gamma _3}\nn\\
&&+\frac{\beta _2
	\left(-2 \beta _2 \beta _3 \beta _1+6 \beta _3 \beta _1-3 \beta _1-3 \beta _3+2\right)}{\gamma _2}\Big),
\nn\\
\Sigma_2 &=&\frac{3 \sqrt{(D-3) (D-2)} \mu  \Omega _{D-2}}{8 \sqrt{10} \pi \gamma _1 \gamma _3 }\left(\beta _1-\beta _3\right) \left(-\beta _1 \beta _3 \beta _2^2+2 \beta _1 \beta _3 \beta _2-2 \beta _2+1\right).
\eea
All the thermodynamic quantities and the scalar charges reduce to those of $B_2$ black hole when $\beta_3=\beta_1$. As we shall see in the next section, the scalar charges play an important role in computing the thermodynamics without the black hole solution, event though they are not thermodynamic quantities themselves.

\section{Black hole thermodynamics without black hole solutions}
\label{sec:thermo}

We have so far considered the construction of two-charge static black holes in the EMMD theories. For suitable dilatonic coupling constants $(a_1,a_2)$, we have seen that the equations of motion can be reduced to Toda equations of all rank-2 Lie groups. This allows us to obtain Toda black holes of Lie groups $D_2$, $A_2$, $B_2$ and $G_2$, and study their black hole thermodynamic properties. However, it was demonstrated \cite{Lu:2025eub,Yang:2025rud} that all the thermodynamic quantities of EMMD black holes can actually be computed without having to know the black hole solution at all. The demonstration was analytically confirmed in four dimensions by all the one-charge EMD black holes and the previously-known two-charge $D_2$, $A_2$ Toda black holes, in addition to some numerical examples. Such a claim was also expected to be true in general $D$ dimensions.

In this section, we generalize the procedure of computing black hole thermodynamic quantities without black hole solution to general $D$ dimensions and test them explicitly with our newly-constructed Toda black holes. As was discussed in \cite{Lu:2025eub,Yang:2025rud}, there are four criteria in the procedure. The first is the weak version of the no-scalar hair conjecture. The general spherically-symemtric and static black holes in the EMMD theory have four independent parameters, $(M,Q_1^e, Q_2^e, \Sigma)$. For the solution to be a black hole, the scalar charge $\Sigma$ can still be turned on, but it cannot be independent, but a specific fine-tuning function of mass and charges, as in \eqref{sigmamq1q2}. A standard procedure of determining this function requires the explicit knowledge of the black hole solution. In order to solve for this relation, we first introduce the net strength $\mu$ of the long-range force of two identical black holes \eqref{forcelaw}. This allows us to express the scalar charge and all the thermodynamic quantities including mass as a function of three basic variables $(Q_1^e, Q_2^e, \mu)$, i.e.
\be
X=X(Q_1^e,Q_2^e,\mu)\,.
\ee
The EMMD theory has no extra dimensionful coupling constants beyond the overall Newton's constant, which we set to be unity. This implies that the scalar charge and all the thermodynamic quantities are homogeneous under the constant scaling according to their dimensions.  The basic variables $(Q_1^e, Q_2^e, \mu)$ have the same dimension of $[L]^{D-3}$.  Thus, if a thermodynamic quantity $X$ has dimension $[L]^n$, we have
\be
X(\lambda Q_1^e, \lambda Q_2^e, \lambda\mu)=\lambda^n X(Q_1^e,Q_2^e,\mu)\,.
\ee
Furthermore, the required fine-tuning of the scalar charge of EMMD black hole must satisfy
\be
\Sigma= \fft{1}{2}a_1 Q_1^e \fft{\partial M}{\partial Q_1^e} +
\fft{1}{2} a_2 Q_2^e \fft{\partial M}{\partial Q_2^e} + \mu\fft{\partial \Sigma}{\partial \mu}\,.
\ee
The above properties allow one to determine the mass $M$ and $\Sigma$ in terms of $(Q_1^e,Q_2^e,\mu)$ explicitly, namely
\be
M=M(Q_1^e, Q_2^e, \mu)\,,\qquad \Sigma=\Sigma(Q_1^e, Q_2^e, \mu)\,.
\ee
Eliminating the parameter $\mu$, we obtain \eqref{sigmamq1q2}.

To determine the thermodynamic quantities associated with the horizon, we need to combine \eqref{forcelaw} and horizon-asymptotic relation \eqref{TSproduct}, which leads to
\be
M^2 + \fft{4}{\delta} \Big(\Sigma^2 -(Q_1^e)^2 +(Q_2^e)^2\Big)  = \fft{4}{\delta^2}(TS)^2\,.\label{forcelaw1}
\ee
Together with the assumption of the first law, we can determine all the remaining thermodynamic quantities such as the temperature, entropy and two electric potentials, without needing to know the exact solution.

Note that the above also involves differential equations and hence we need some ``initial condition'' to solve for all the thermodynamic quantities directly without appealing to the black hole solutions. It was shown in \cite{Lu:2025eub,Yang:2025rud} that the only input needed is the Schwarzschild black hole thermodynamics and we may treat them as initial conditions. The first step is to use the Schwarzschild black hole thermodynamics as initial condition and derive the one-charge black hole thermodynamics, which turns out to be analytic. The general two-charge thermodynamics can then be computed in the second step, using the obtained one-charge results as the initial conditions. The procedures were given in great detail in \cite{Lu:2025eub,Yang:2025rud} and we shall not repeat them here.

It is easy to verify that the black hole thermodynamic quantities and the scalar charge in our Toda black holes, given explicitly in previous sections, satisfy all the criteria listed above. We therefore confirm the validity of the procedure of computing black hole thermodynamics without black hole solution with two more concrete and nontrivial examples. Guided by this success, we may further generalize the EMMD theory and consider Einstein gravities minimally coupled to $n$ Maxwell fields and $m\le n$ dilaton scalars.  The Lagrangian takes the form \cite{Lu:2025eub}
\be
{\cal L}=\sqrt{-g} \Big(R - \fft12(\partial \vec \phi)^2 - \fft14 \sum_{i=1}^n e^{\vec a_i \cdot \vec \phi} (F_i)^2\Big)\,,\qquad  F_i = dA_i\,,\qquad \vec \phi=(\phi_1,\phi_2,\cdots,\phi_m)\,.
\ee
We refer to these theories with multiple Maxwell and dilaton fields as EMDS, where the last $S$ in the acronym symbolizes the plural of the Maxwell and dilaton fields. The multi-charge black holes now have mass $M$, electric charges $Q_i^e$ and scalar charges $\vec \Sigma = (\Sigma_1, \ldots, \Sigma_m)$. It is reasonable to conjecture that we can also compute the thermodynamics of the spherically-symmetric and static black holes in EMDS theories without having to know the solutions. Based on the EMMD experience, the two key equations are clearly
\bea
&& M^2 + \fft{4}{\delta} \Big(\vec \Sigma \cdot \vec \Sigma- \sum_{i} (Q_i^e)^2\Big)  = \fft{4}{\delta^2}(TS)^2\,,\nn\\
&&\vec\Sigma=\mu\fft{\partial \vec \Sigma}{\partial \mu}+ \fft12 \sum_{i=1} \vec a_i\, Q_i^e \fft{\partial M}{\partial Q_i^e} \,.
\eea
It is easy to verify that the thermodynamics of STU black holes in four and five dimensions indeed satisfy these equations. (The relevant black hole solutions can be found in \cite{Behrndt:1998jd,Duff:1999gh,Cvetic:1999xp}, but the cosmological constant should be turned off for our purpose.) More nontrivially, in section \ref{sec:a3b2}, we presented $A_3$ Toda black holes, involving three Maxwell fields and two dilatons. It can be verified that its thermodynamic quantities and the scalar charges also satisfy the above two key equations. Thus, for these theories, the black hole thermodynamics can indeed be computed without needing to know the black hole solution. Our work has therefore increased the repertoire of the theories that allow to a direct computation of black hole thermodynamics without solutions.

\section{Conclusions}
\label{sec:con}

In this paper, we completed two tasks. The first task was the construction of Toda solutions of rank-2 Lie groups and associated exact black holes. We studied the construction of spherically-symmetric and static black holes charged under two maxwell fields in EMMD theories in general $D$ dimensions. We first cast the equations of motion into a set of Toda-like equations and they become integrable Toda equations of all rank-2 Lie groups for suitable dilaton coupling constants. We devised a brute-force method and obtained the most general and elegant solutions to the Toda equations. The general solutions, which involve naked singularities, have four independent parameters, namely the mass, scalar charge and two electric charges. Black hole solutions with regular event horizon emerge if we fine-tune the scalar charge to some specific function of mass and electric charges.

The second task involved the study of the black hole thermodynamics and we used the two new black holes to further confirm the claim of \cite{Lu:2025eub,Yang:2025rud} that the thermodynamics of the EMMD black holes can be obtained without having to know the black hole solutions. We then generalized the claim further to include EMDS theories, where multiple Maxwell and dilaton fields are involved. We used the known $A_3$ Toda black hole to confirm the claim. It is rather intriguing to note that although the scalar charges are not themselves thermodynamic quantities that enter the first law, they play a vital role in computing the thermodynamics directly without solutions. Our work expands the portfolio of theories where black hole thermodynamics can be computed without black hole solutions.

In this paper, we considered only the black holes that are asymptotic to Minkowski spacetimes. For suitable scalar potentials, the $D_2$ and $A_2$ Toda black holes were generalized to be asymptotic AdS \cite{Lu:2013eoa,Lu:2013ura}. It is of great interest to investigate whether such generalizations are possible for general Toda black holes.

\section*{Acknowledgement}

We are grateful to Hai-Shan Liu and Pujian Mao for useful discussions. This work is supported in part by the National Natural Science Foundation of China (NSFC) grants No.~12375052 and No.~11935009, and also by the Tianjin University Self-Innovation Fund Extreme Basic Research Project Grant No.~2025XJ21-0007.

\appendix

\section{Review of $D_2$ and $A_2$ Toda black holes}
\label{app:review}

In the main text, we construct spherically-symmetric and static black holes charged under both Maxwell fields in the EMMD theory. We showed that the equations of motion can be cast into two Toda-like equations. For suitable dilaton couplings, they are Toda equations associated with Lie groups of rank 2. The $D_2$ and $A_2$ Toda black holes are known in the literature and for the self containment of this paper, we list them in this appendix.

\subsection{$D_2$ Toda black holes}

The $D_2\sim A_1\times A_1 $ black hole arises when the dilaton coupling constants satisfy $a_1 a_2=-\delta$. The most general two-charge non-extremal black hole solutions are known in literature, given by \cite{Lu:2013eoa}
\be
H_1 = 1 + \fft{p_1}{r^{D-3}}\,,
\qquad
H_2  = 1 + \fft{p_2}{r^{D-3}}\,,
\ee
where the two constants are related to the electric charge parameters $(Q_1,Q_2)$ as follows
\be
Q_1=\frac{\delta  \sqrt{p_1 \left(2 \mu +p_1\right)}}{2 (2-\delta ) \sqrt{\delta +a_1^2}}\,,\qquad
Q_2=\frac{\delta  \sqrt{p_2 \left(2 \mu +p_2\right)}}{2 (2-\delta
	) \sqrt{\delta +a_2^2}}\,.
\ee
We can read off the electric charges from \eqref{echarges}. The electric potentials can be obtained from integrating the Maxwell field strengths $F_1$ and $F_2$, given by
\be
\Phi_1=2 \sqrt{\frac{r_+^{3-D} p_1}{\left(\delta +a_1^2\right) H_1\left(r_+\right)}}\,,\quad
\Phi_2= \sqrt{\frac{r_+^{3-D} p_2}{\left(\delta +a_2^2\right)
		H_2\left(r_+\right)}}
\ee
The mass and the scalar charge are given by
\bea
M &=&  \fft{(D-2)\Omega_{D-2}}{8\pi}\Big(\mu+ \fft{\delta}{\delta + a_1^2} q_1 +
\fft{\delta}{\delta + a_2^2} q_2 \Big)\,,\nn\\
\Sigma &=&  \sqrt{\fft{(D-3)(D-2)}{32 \pi^2}}\Omega_{D-2} \Big(\fft{a_1}{\delta +a_1^2} p_1+\fft{a_2}{\delta +a_2^2} p_2\Big).
\eea
The temperature and entropy are given by \eqref{TandS}. We can then easily verify that the first law \eqref{firstlaw} holds. The solution reduces to the RN black hole with $p_1=p_2$. Note that if we use $(\beta_1,\beta_2, \mu)$ parameters, the solution becomes
\be
H_i = \fft{1}{1-\beta_i} \big(1 - \beta_1 f)\,,\qquad \hbox{with}\qquad
p_i = \fft{2\mu\beta_i}{1-\beta_i}\,,\qquad i=1,2.
\ee

\subsection{$A_2$ Toda black holes}

The $A_2$ Toda black hole was first constructed in \cite{Gibbons:1985ac}, as the dyonic black hole of the Kaluza-Klein theory. For our EMMD theory, it arises when then the dilaton coupling constants are $a_1=-a_2=\sqrt{3\delta}$. Adopting the notation of \cite{Lu:2013uia}, the general non-extremal two-charge black hole is given by
\be
H_1 = \fft{1}{\gamma_1} (1 - 2 \beta_1 f + \beta_1\beta_2 f^2)\,,\qquad
H_2 = \fft{1}{\gamma_2} (1 - 2 \beta_2 f + \beta_1\beta_2 f^2)\,,
\ee
with $\gamma_1 = 1 -2 \beta_1 + \beta_1\beta_2$ and $\gamma_2 = 1 - 2\beta_2 + \beta_1\beta_2$. The equations of motion are all satisfied provided that
\be
Q_1^e=\fft{\Omega_{D-2}}{8\pi} \sqrt{(D-3)(D-2)\beta_1 \gamma_2}\, \gamma_1^{-1} \mu\,,\qquad
Q_2^e=\fft{\Omega_{D-2}}{8\pi} \sqrt{(D-3)(D-2)\beta_2 \gamma_1}\, \gamma_2^{-1} \mu\,.
\ee
The corresponding electric potentials are
\be
\Phi_1= \fft{\sqrt{2\beta_1} (1-\beta_2)}{\sqrt{\delta\, \gamma_2}}\,,\qquad
\Phi_2= \fft{\sqrt{2\beta_2} (1-\beta_1)}{\sqrt{\delta\, \gamma_1}}\,.
\ee
The mass and the scalar charge is
\bea
M &=& \fft{(D-2)(1-\beta_1)(1-\beta_2)(1-\beta_1\beta_2) \Omega_{D-2}}{
8\pi \gamma_1 \gamma_2}\,,\nn\\
\Sigma &=& \fft{\sqrt3 (D-2) (\beta_1-\beta_2)(1-\beta_1\beta_2) \Omega_{D-2}}{8\pi
\gamma_1\gamma_2}\,.
\eea
The temperature and entropy are given by \eqref{TandS}. It is easy to verify that the first law \eqref{firstlaw} holds. The solution reduces to the RN black hole when $\beta_2=\beta_1$.

\section{Extremal $B_2$, $G_2$ black holes in explicit charges}
\label{app:extremal}

In section \ref{sec:b2g2}, we obtained exact $B_2$ and $G_2$ Toda black holes. The general non-extremal solutions involves three parameters, the mass and two electric charges, with all thermodynamic quantities expressed in terms of parametric functions of $(\beta_1, \beta_2, \mu)$. In the extremal limit, mass and all other thermodynamic quantities are expressed in terms of $(c_1,c_2)$, parametrizing the two electric charges. In this appendix, we present the extremal solutions in terms of electric charges directly. For simplicity, we shall first give the $D=4$ case, where $(Q_1^e, Q_2^e)=(Q_1,Q_2)$. The extremal solution, parametrized in $(Q_1,Q_2)$, is given by
\bea
H_1 &=& 1 -\frac{2\sqrt{10 Q_1} \left(\sqrt{Q_1}-2 c^3 \sqrt{Q_2}\right)}{r}+\frac{40 c^2 Q_1 Q_2}{r^2}+\frac{80 \sqrt{10} c Q_1^{3/2} Q_2^{3/2}}{3 r^3}+\frac{200 Q_1^2 Q_2^2}{3 r^4}\,,\nn\\
H_2 &=& 1+\frac{2 \sqrt{10} c^2 Q_2}{r}+\frac{20 c \sqrt{Q_1} Q_2^{3/2}}{r^2}+\frac{20 \sqrt{10} Q_1 Q_2^2}{3 r^3}\,,
\eea
where the constant parameter $c$ satisfies a quartic equation
\be
\left(1-2 c^4\right) Q_2+2 c \sqrt{Q_1 Q_2}=0\,.\label{ccons}
\ee
The mass and the scalar charge are
\be
M=\sqrt{\frac{2}{5}} \left(2 c^3 \sqrt{Q_1 Q_2}+2 c^2 Q_2-Q_1\right)\,,\qquad
\Sigma = \sqrt{\frac{1}{10}} \left(6 c^3 \sqrt{Q_1 Q_2}-4 c^2 Q_2-3 Q_1\right)\,.
\ee
It is easy to verify that the long-range force vanishes and we also have
\be
\Sigma = \frac12 a_1 Q_1 \fft{\partial M}{\partial Q_1} +  \frac12 a_2 Q_2 \fft{\partial M}{\partial Q_2}\,.\label{sigmaext}
\ee
A consistency check is whether the extremal solution can reduce to RN extremal black hole. The condition is take
\be
Q_1 = \sqrt{\fft{-a_2}{a_1-a_2}} Q\,,\qquad Q_2 = \sqrt{\fft{a_1}{a_1-a_2}} Q\,.
\ee
This implies that
\be
Q_1= \sqrt{\ft25} Q\,,\qquad Q_2 = \sqrt{\ft35} Q\,.
\ee
The constraint \eqref{ccons} can be solved with $c=\big(\ft32\big)^{\fft14}$. The solution now reduces to the RN extremal black hole, given by
\be
H_1 = H^4\,,\qquad H_2=H^3\,,\qquad \hbox{with}\qquad H=1 + \fft{2Q}{r}\,.
\ee
In this limit, we have $M=2Q$, precisely the mass/charge relation of the RN.

The extremal $G_2$ Toda black hole solution in $D=4$ is given by
\bea
H_1 &=&1 + \frac{\sqrt{\frac{7}{3}} c^3 Q_1^2 \left(c^6 Q_1^2-72 Q_2^2\right)}{24 Q_2^3 r} +\frac{7 c^2 Q_1^2 \left(c^6 Q_1^2-48 Q_2^2\right)}{6 Q_2^2 r^2} +\frac{28 \sqrt{\frac{7}{3}} c Q_1^2 \left(c^6 Q_1^2-24 Q_2^2\right)}{3 Q_2 r^3}\nn\\
&&+\frac{392 Q_1^2 \left(c^6 Q_1^2-8 Q_2^2\right)}{3 r^4} + \frac{1568 \sqrt{\frac{7}{3}} Q_1^2 Q_2 \left(17 c^6 Q_1^2-24 Q_2^2\right)}{45 c r^5} + \frac{614656 c^4 Q_1^4 Q_2^2}{135 r^6}\nn\\
&&+\frac{1404928 \sqrt{\frac{7}{3}} c^3 Q_1^4 Q_2^3}{135 r^7}+\frac{4917248 c^2 Q_1^4 Q_2^4}{135 r^8}+\frac{39337984 \sqrt{\frac{7}{3}} c Q_1^4 Q_2^5}{1215 r^9}\nn\\
&&+\frac{550731776 Q_1^4 Q_2^6}{18225 r^{10}}\,,\nn\\
H_2 &=& 1  +\frac{\sqrt{\frac{7}{3}} \left(c^6 Q_1^2+24 Q_2^2\right)}{6 c Q_2 r} +
\frac{14 c^4 Q_1^2}{3 r^2}+\frac{224 \sqrt{\frac{7}{3}} c^3 Q_1^2 Q_2}{9 r^3}+\frac{1568 c^2 Q_1^2 Q_2^2}{9 r^4}\nn\\
&&+\frac{12544 \sqrt{\frac{7}{3}} c Q_1^2 Q_2^3}{45 r^5}+\frac{175616 Q_1^2 Q_2^4}{405 r^6}\,,
\eea
where the dimensionless constant $c$ satisfies the constraint
\be
576 (c^2 - 1)  - c^6 \Big( c^6 \fft{Q_1^4}{Q_2^4} - 96 \fft{Q_1^2}{Q_2^2}\Big)=0\,.\label{ccons2}
\ee
The mass and scalar charge in this extremal case are
\bea
M &=& \frac{c^{10} Q_1^4+12 \left(c^2-6\right) c^4 Q_1^2 Q_2^2+288 Q_2^4}{96 \sqrt{21} c Q_2^3}\,,\nn\\
\Sigma &=& \frac{3 c^{10} Q_1^4-4 \left(5 c^2+54\right) c^4 Q_1^2 Q_2^2-480 Q_2^4}{96 \sqrt{7} c Q_2^3}\,.
\eea
It is easy to verify that the long-range force indeed vanishes in the extremal limit, and furthermore the relation \eqref{sigmaext} holds. To reduce the solution to RN, we must have
\be
Q_1=\sqrt{\ft5{14}} Q\,,\qquad Q_2 = \ft{3}{\sqrt{14}} Q\,.
\ee
In this case, the constraint can be solved with $c=\sqrt6$. The mass is then again simply given by $M=2Q$, with the solution given by
\be
H_1 = H^{10}\,,\qquad H_2= H^{6}\,,\qquad H = 1 + \fft{2Q}{r}\,.
\ee

In the above, we present the extremal $B_2$ and $G_2$ Toda black holes in four dimensions. For general $D$ dimensions, we only need to make the following replacement, namely
\be
r\rightarrow r^{D-3}\,,\qquad Q_i \rightarrow \sqrt{\fft{2}{(D-3)(D-2)}} Q_i\,,
\ee
where the redefined two $Q_i$'s are those appearing in ansatz \eqref{ansatz-1}. Note that the equations \eqref{ccons} and \eqref{ccons2} for the dimensionless parameter $c$ remain unchanged.

\end{document}